\documentclass[prd,
tightenlines,nofootinbib,
  eqsecnum
  ]{revtex4}

\usepackage{amsmath}
\usepackage{amsfonts}
\usepackage{amssymb}
\usepackage{bm}
\usepackage{hyperref}
\usepackage{mathrsfs}
\usepackage{graphicx}

\usepackage{ulem}
\usepackage[usenames]{color}

\definecolor{darkgreen}{rgb}{0,0.5,0}

\hypersetup{
    bookmarks=true,         
    unicode=false,          
    pdftoolbar=true,        
    pdfmenubar=true,        
    pdffitwindow=false,     
    pdfstartview={FitH},    
    pdftitle={My title},    
    pdfauthor={Author},     
    pdfsubject={Subject},   
    pdfcreator={Creator},   
    pdfproducer={Producer}, 
    pdfkeywords={keyword1} {key2} {key3}, 
    pdfnewwindow=true,      
    colorlinks=true,       
    linkcolor=red,          
    citecolor=cyan,        
    filecolor=magenta,      
    urlcolor=darkgreen,           
    linktocpage=true
}

\allowdisplaybreaks

\DeclareSymbolFontAlphabet{\mathrsfs}{rsfs}
\DeclareMathAlphabet{\mathcal}{OMS}{cmsy}{m}{n}

\newcommand{\ud}{\mathrm{d}}

\newcommand{\beq}{\begin{equation}}
\newcommand{\eeq}{\end{equation}}

\newcounter{theorem} 

\begin{document}

\title{Analyzing Gravitational Waves with General Relativity\footnote{Invited review article to appear in Comptes Rendus de l'Acad\'emie des Sciences (Physique).}}

\author{Luc \textsc{Blanchet}}\email{luc.blanchet@iap.fr}
\affiliation{
  Institut d'Astrophysique de Paris,\\ UMR 7095, CNRS, Sorbonne
  Universit{\'e},\\ 98\textsuperscript{bis} boulevard Arago, 75014 Paris,
  France}

\date{\today}

\begin{abstract}
\noindent\textit{English:} After a short review of prominent properties of gravitational waves and the newly born gravitational astronomy, we focus on theoretical aspects. Analytic approximation methods in general relativity have played a crucial role in the recent discoveries of gravitational waves. They are used to build theoretical template banks for searching and analyzing the signals in the ground-based detectors LIGO and Virgo, and, further ahead, space-based LISA-like detectors. In particular, the post-Newtonian approximation describes with high accuracy the early inspiral of compact binary systems, made of black holes or neutron stars. It mainly consists of extending the Einstein quadrupole formula by a series of relativistic corrections up to high order. The compact objects are modelled by point masses with spins. The practical calculations face difficult problems of divergences, which have been solved thanks to the dimensional regularization. In the last rotations before the merger, the finite size effects and the internal structure of neutron stars (notably the internal equation of state) affect the evolution of the orbit and the emission of gravitational waves. We describe these effects within a simple Newtonian model.

\smallskip
\noindent\textit{Fran\c{c}ais:} Apr\`es une br\`eve revue des propri\'et\'es importantes des ondes gravitationnelles et de la nouvelle astronomie gravitationnelle, nous nous concentrons sur les aspects th\'eoriques. Les m\'ethodes d'approximation analytiques en relativit\'e g\'en\'erale ont jou\'e un r\^ole crucial dans les r\'ecentes d\'etections d'ondes gravitationnelles. Elles sont utilis\'ees pour cr\'eer des banques de mod\`eles (patrons) th\'eoriques qui servent \`a rechercher et analyser les signaux dans les d\'etecteurs au sol LIGO et Virgo et, plus tard, les d\'etecteurs dans l'espace de type LISA. En particulier, l'approximation post-newtonienne d\'ecrit avec grande pr\'ecision le spiralement initial des syst\`emes binaires compacts de trous noirs ou d'\'etoiles \`a neutrons. Elle consiste principalement \`a \'etendre la formule du quadrup\^ole d'Einstein par une s\'erie de corrections relativistes jusqu'\`a un ordre \'elev\'e. Les objets compacts sont mod\'elis\'es par des masses ponctuelles avec spins. Les calculs pratiques font face \`a des probl\`emes difficiles de divergences, qui ont \'et\'e r\'esolus gr\^ace \`a la r\'egularisation dimensionnelle. Dans les derni\`eres orbites proches de la fusion, les effets de taille finie et de structure interne des \'etoiles \`a neutrons (notamment l'\'equation d'\'etat interne) affectent l'\'evolution de l'orbite et l'\'emission des ondes gravitationnelles. Nous d\'ecrivons ces effets dans le cadre d'un mod\`ele newtonien simple.

\smallskip
\noindent\textit{keywords:} Gravitational waves; compact binary systems; post-Newtonian theory.
\end{abstract}


\maketitle

\section{Gravitational waves and the new astronomy}
\label{sec:gravastro}

Paramount breakthroughs in Astronomy and fundamental Physics occurred with the discovery of gravitational waves (GWs) generated by the orbital motion and merger of compact binary systems, made of black holes or neutron stars~\cite{GW150914,GW170817}. A new window of observations of our Universe opened up, radically different and complementary from that of the traditional astronomy, essentially based on electromagnetic (EM) waves. The salient properties of GWs shape the key features of the new ``Gravitational Astronomy'': 
\begin{itemize}
\item GWs are produced by the overall, ``bulk'' motion of large masses at relativistic speeds (close to the speed of light $c$), in contrast to EM waves which are in general composed of the incoherent superposition of photons emitted by the atoms and molecules composing the source~\cite{Th300}. As a result the wavelength of GWs is in general much larger than the size of the source, and there is a deep analogy between GWs and ordinary sound waves. However, in contrast to sound waves, GWs propagate in vacuum. They are ripples in the Riemannian curvature of space-time, which is the fundamental dynamical entity in general relativity (GR).
\item GWs propagate almost without alteration through the densest regions of the Universe, and have thus the potential of carrying information from very far away --- probably up to the first instants after the Big Bang. This is due to the weakness of the gravitational interaction as compared to other forces, and to the fact that GWs cannot be screened by any type of matter field. Indeed the charge associated with the gravitational interaction is the mass, which is always equal to the inertial mass by the equivalence principle, and therefore is always positive. Actually the positivity of the mass-energy of an arbitrary system (involving ordinary bodies and black holes), constitutes an important and difficult theorem in GR~\cite{SchoenYau,Witten}. 
\item GWs emitted by coalescing compact binary systems contain the information about their distance~\cite{Schutz86}. In this respect these systems constitute an analogue of the standard candels of EM-based astronomy (like Cepheid variables and Type Ia supernovas), and can rightly be called ``\textit{standard sirens}''. However in the case of the GW sirens there is no need for calibrating the distance scale; the calibration is automatically done by GR. One can thus measure the Hubble-Lema\^itre cosmological parameter $H_0$ with GWs, independently of the traditional EM measurements.\,\footnote{A resolution of the XXX$^\text{th}$ general assembly of the international astronomical union (IAU) recommended that the expansion of the Universe be referred to as the ``Hubble-Lema\^itre law''.}
\item The gravitational astronomy is one of ``\textit{precision}''. It is possible to measure with high precision the parameters of compact binary systems (masses and spins) by direct comparison with a solution of the purely gravitational two-body problem in GR. See Fig.~\ref{fig1} which shows the signal of the first binary black hole event, directly analyzed with the GR prediction. So far no deviation from GR has been observed. For compact sources most of the non gravitational effects, which usually plague the dynamics of ordinary systems (magnetic fields, presence of an interstellar medium, \textit{etc.}), are in general dominated by the gravitational force.\,\footnote{We shall discuss an exception in Sec.~\ref{sec:finitesize}: The internal structure and the non-gravitational equation of state of neutron stars do affect the GW signal close to and during the final merger.}  
\begin{figure}[t]
\begin{center}
\includegraphics[width=13cm]{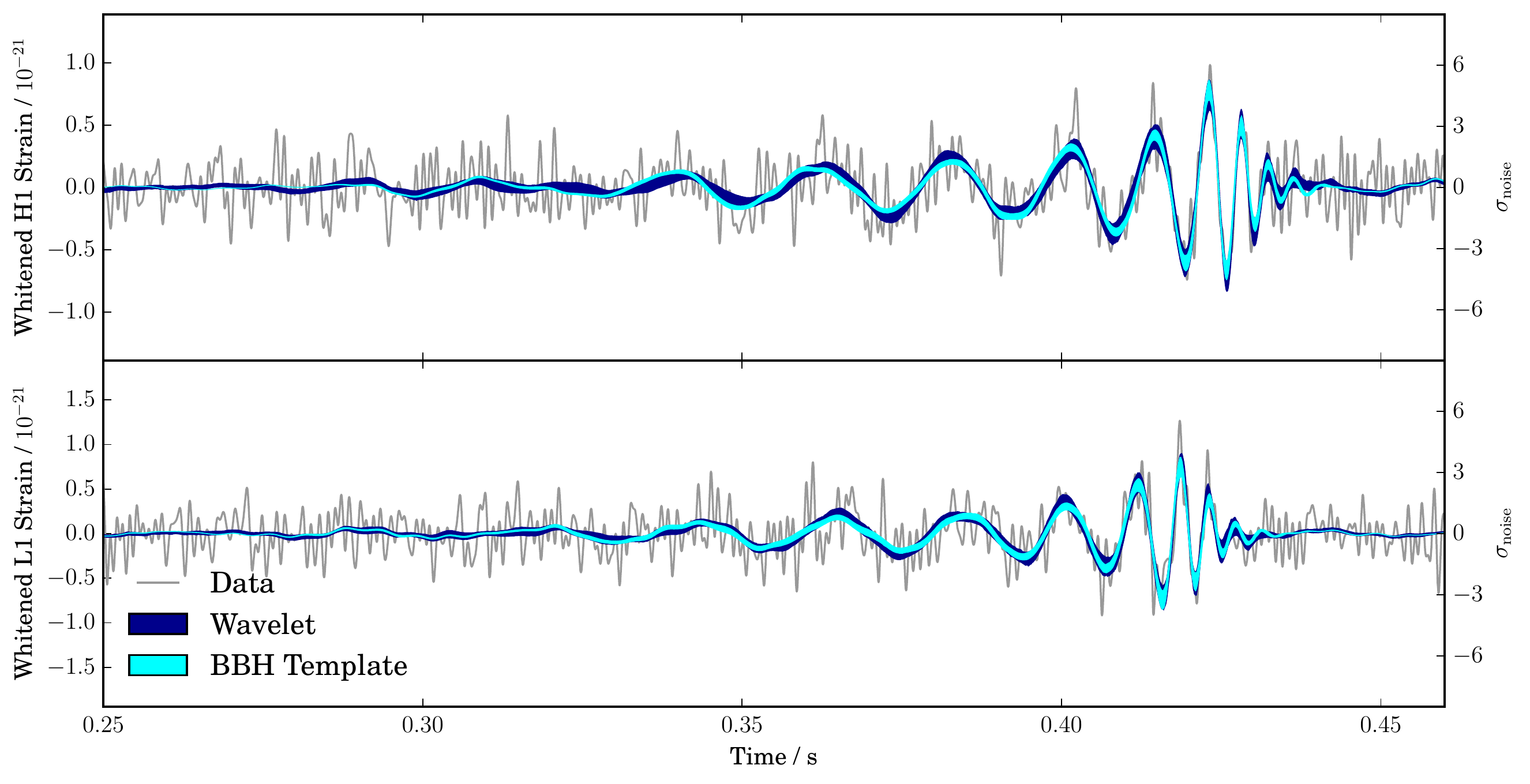}
\end{center}
\caption{The first binary black hole event of September 14, 2015 (GW150914) seen by the two LIGO detectors~\cite{GW150914}. The signal can be directly confronted with the GR prediction. Shown is the best adjustment with the result from the modeled analyses using IMR-Phenom and EOB-NR template waveforms (see Sec.~\ref{sec:analapprox}). These analyses are in agreement with the full numerical calculation of the merger of two black holes. The last cycles before the merger are also reasonably well interpreted by the quadrupole formula. The masses and spins of the black holes are inferred from the comparison with GR. The signal can also be matched by a superposition of wavelets, but these are devoided of any physical content.}\label{fig1}
\end{figure}
\item This highlights the crucial role played by analytic approximation methods (reviewed in Secs.~\ref{sec:quadformula} and~\ref{sec:analapprox}) and also numerical calculations, since they permit an accurate description of the two-body problem in GR, without which the full information contained into the signal could not be extracted. The theoretical solution of the problem of motion and radiation is used in the form of accurate GW templates, which are correlated to the observed signal using the technique of matched filtering~\cite{BuonSathya15}.
\item The new astronomy is also ``\textit{fundamental}''. As far as we know, our gravitational theory is fundamental, and GR may be valid in a large range of energies, perhaps up to the Planck scale. Thus GW observations have a lot of implications for fundamental physics. With GWs one can confront GR with alternative theories of gravity (such as scalar-tensor theory, massive gravity theory, \textit{etc.}) and one can test fundamental principles such as the equivalence principle. One can also question the standard model of cosmology $\Lambda$-CDM, including the great mysteries of contemporary physics constituted by cold dark matter (CDM), the cosmological constant $\Lambda$ and dark energy~\cite{roadmap}.
\item Last but not least: The multi-messenger aspect of the gravitational astronomy, \textit{i.e.}, its synergy with other vectors of information, most importantly EM waves but maybe also neutrinos in the future. For instance, the joint observation of GWs from a binary neutron star event and of a gamma ray burst (GRB) permitted to show that the speed of GWs is equal to $c$ with a precision $\sim 10^{-15}$. This ruled out a series of alternative theories of gravity. But of course, the multi-messenger astronomy has outstanding implications in astrophysics, such as refining the model of GRBs, understanding the explosions of kilonovas and the mechanisms for the production of heavy elements.
\end{itemize}

Consider an isolated (with finite spatial support) source of GWs. Let $m$ be the mass of the source, $r$ its size and $\omega \sim 2\pi/P$ the typical angular frequency of oscillations of the source, with $P$ the typical period. We have $\omega \sim v/r$ with $v$ the typical internal velocity. We suppose that the source is self gravitating, so the gravitational force is responsible for the dynamics and the GW generation. The wavelength of the emitted GW is $\lambda \sim c P/2$, with a factor $1/2$ inserted to take into account the quadrupolar nature of GWs, whose frequency is $f \sim \omega/\pi$. Note that $r/\lambdabar \sim 2v/c$ (posing $\lambda=2\pi\lambdabar$), so that the source is much smaller than one wavelength of the emitted GW, since $\epsilon_\text{PN}\sim v/c \ll 1$ in the non relativistic (post-Newtonian) regime. For the self gravitating source we typically have $G m \sim r^3\omega^2$, indeed this is exactly Kepler's third law $G m = a^3 \omega^2$ in the case of a Newtonian binary system (with $a$ the semi-major axis of the orbit). Furthermore, if the source is \textit{compact} its size is of the order of the Schwarzschild radius, $r \sim 2 G m/c^2$, and the GW frequency scales inversely proportional to the mass: $G m f \sim c^3/(\pi\sqrt{8})$. But the mass decreases because the GW extracts energy from the source, hence the frequency of the GW increases: This is the famous ``chirp'' of GWs, which we shall compute with high precision in Sec.~\ref{sec:analapprox}.

\section{Successes with the Einstein quadrupole formula}
\label{sec:quadformula}

The ``precision'' gravitational astronomy requires inputs from the theory side. Searching and analyzing GW signals that are well predicted by GR is made using the technique of matched filtering, which cross correlates the detector output with our best prediction of the expected signal, called the template. The template is weighted (in the Fourier domain) by the power spectral density of the noise in the detector. It depends on a set of trial parameters describing the source's model (such as masses and spins), and that are measured in the process. As GR is a complicated non-linear theory, there is no hope of finding an exact solution of the Einstein field equations, but tremendous progresses have been made with the development of perturbative and approximation methods in GR, notably the post-Newtonian (PN) approximation. Conjointly with analytic developments, continued efforts in numerical relativity led to the computation of the final merger of binary black holes and other GW sources like supernova explosions.

The first analytic computation of GWs is the Einstein quadrupole formula~\cite{E16,E18}, valid at the dominant ``Newtonian'' order in a PN expansion, with the small PN parameter being the slowness estimate $\epsilon_\text{PN}\sim v/c$, ratio of a typical internal velocity in the source and the speed of light. Originally derived for matter sources with negligible self gravity (hence the source's oscillations producing GWs have a non gravitational origin), the formula was later shown to be still valid for weakly self-gravitating sources, such as a Newtonian binary system~\cite{LL}. The GW amplitude is characterized by two tensorial polarization modes, traditionally denoted $h_+$ and $h_\times$, that are transverse to the direction of propagation $\bm{n}=(n^i)$ (with $i=1,2,3$ and $\bm{n}^2=1$), pointing from the GW source towards a far-away detector. The detector is sensitive to a certain linear combination of the two polarizations, 
\begin{equation}\label{h}
h=\mathcal{F}_+ h_+ + \mathcal{F}_\times h_\times\,,
\end{equation}
where the ``form factors'' $\mathcal{F}_{+}$ and $\mathcal{F}_{\times}$ depend on the direction and orientation of the source with respect to the local frame of the detector. The polarizations are defined as the projection of the waveform along two polarization vectors $\bm{p}$ and $\bm{q}$ in the plane orthogonal to $\bm{n}$, and forming with it an orthonormal right-handed triad $(\bm{n},\bm{p},\bm{q})$. The quadrupole formula gives the polarizations at large distance $d$ from the source (and at retarded time $t-d/c$) as
\begin{equation}\label{polar}
\left(\begin{array}{l}h_+\\[0.2cm]h_\times
\end{array}\right) = \frac{2G}{c^4 d}
\left(\begin{array}{l}\frac{p^i p^j - q^i q^j}{2}\\[0.2cm]\frac{p^i q^j + p^j q^i}{2}
\end{array}\right)\biggl\{\frac{\ud^2 Q_{ij}}{\ud t^2}\bigl(t-d/c\bigr)+
  \mathcal{O}\left(\epsilon_\text{PN}\right)\biggr\} +
  \mathcal{O}\left(\frac{1}{d^2}\right) \,.
\end{equation}
The quadrupole moment $Q_{ij}$ of the source is just, at the leading approximation, the usual \textit{mass type} moment of the Newtonian mass density $\rho$ in the source,
\begin{equation}\label{Qij}
Q_{ij} = \int \ud^3\mathbf{x} \,\rho\,\Bigl(x^i x^j - \frac{1}{3} \mathbf{x}^2 \Bigr) \,.
\end{equation}
The total energy $E$ of the matter source decreases because of the GW emission, and this is controlled by the ``flux-balance'' equation
\begin{equation}\label{balanceE}
\frac{\ud E}{\ud t} = - F^\text{GW} \,,
\end{equation}
where the GW flux in the right-hand side is given at the leading PN approximation (\textit{i.e.}, ``Newtonian'' order in the radiation field) by
\begin{equation}\label{fluxE}
F^\text{GW} \equiv \frac{G}{5 c^5} \left\{\frac{\ud^3 Q_{ij}}{\ud t^3}\frac{\ud^3 Q_{ij}}{\ud t^3} 
  + \mathcal{O}\left( \epsilon_\text{PN}^2\right)\right\}\,.
\end{equation}
Witness the factor $c^{-5}$ in front of the flux, which shows that the corresponding radiation reaction effect in the matter equations of motion is actually of order 2.5PN, namely $\mathcal{O}(\epsilon_\text{PN}^5)$.\footnote{By order $n$PN we refer to a small post-Newtonian term of the order of $\epsilon_\text{PN}^{2n}$.} Similarly there is a quadrupole flux-balance equation for the angular momentum $J_i$, given by 
\begin{equation}\label{balanceJ}
\frac{\ud J_i}{\ud t} = - \frac{2G}{5c^5} \, \varepsilon_{ijk} \left\{\frac{\ud^2 Q_{jl}}{\ud t^2}\frac{\ud^3 Q_{kl}}{\ud t^3} +
\mathcal{O}\left( \epsilon_\text{PN}^2\right)\right\} \,.
\end{equation}

The laws of motion of a relativistic conservative system (neglecting the GW emission) admit ten Noetherian invariants associated with the symmetries of the Poincar\'e group. In addition to the energy $E$ and angular momentum $J_i$, there is the linear momentum $P_i$ and the invariant of the center of mass $G_i$. The latter CM invariant is associated with the invariance under Lorentz boosts. When the GWs are turned on, all the invariants obey some flux-balance equations. In addition to~\eqref{balanceE}--\eqref{balanceJ} we have
\begin{subequations}\label{balancePG}
\begin{align}
\frac{\ud P_i}{\ud t} &= - \frac{G}{c^7} \left\{ \frac{2}{63}
  \frac{\ud^4 Q_{ijk}}{\ud t^4}\frac{\ud^3 Q_{jk}}{\ud t^3} + \frac{16}{45}
  \varepsilon_{ijk} \frac{\ud^3 Q_{jl}}{\ud t^3}\frac{\ud^3 D_{kl}}{\ud t^3} +
\mathcal{O}\left( \epsilon_\text{PN}^2\right) 
\right\} \,,\\
\frac{\ud G_i}{\ud t} &= P_i - \frac{G}{c^7} \left\{ \frac{2}{21} \frac{\ud^3 Q_{ijk}}{\ud t^3}\frac{\ud^3 Q_{jk}}{\ud t^3} +
\mathcal{O}\left( \epsilon_\text{PN}^2\right) \right\}\,,\label{balanceG}
\end{align}
\end{subequations}
where $Q_{ijk}$ is the Newtonian mass octupole moment and $D_{ij}$ is the \textit{current type} quadrupole moment. Notice that~\eqref{balancePG} represent subdominant radiation reaction effects of order 3.5PN. The flux of linear momentum $P_i$ is well known, as it is responsible for the ``gravitational recoil'' of the source by GW emission (see, \textit{e.g.},~\cite{BQW05}). However, strangely enough, the expression of the flux of center-of-mass position $G_i$ in Eq.~\eqref{balanceG}, has only been computed and recognized recently~\cite{KNQ18,N18,BF18}.

The equations~\eqref{balanceE}--\eqref{balanceJ} for energy and angular momentum give the evolution of the orbital parameters (semi-major axis and eccentricity) of the compact binary system under GW emission~\cite{PM63,Peters64}. An average over the orbital period is applied, so as to consider the \textit{secular} evolution of the orbit on a radiation reaction time scale much longer than the orbital period. The first success of the quadrupole formula has been that it works perfectly when accounting for the observed decay of the orbital period of the Hulse-Taylor binary pulsar~\cite{HulseTaylor}. This test represented the first quantitatively precise proof of the existence of GWs~\cite{TFMc79,TW82,DT91}. Nevertheless, since $\epsilon_\text{PN}\sim 10^{-3}$ is very small for binary pulsars, the quadrupole formula is not expected to yield any deviation with respect to observations in the regime of binary pulsars.\footnote{But see the controversial debate on this point at the time of the binary pulsar~\cite{Ehletal76,WalkW80}.}

Even more impressive, a second success occurred recently, because the GW signals from binary black hole and neutron star events can be reasonably well interpreted with the quadrupole formula. In the case of black hole binaries this is truly remarkable because $\epsilon_\text{PN}\sim 0.5$ in the last rotations. Take the example of the first event GW150914, shown in Fig.~\ref{fig1}. We are observing the signal at high frequency, close to the final merger, so the orbit has been circularized by radiation reaction --- a consequence of the balance equation for angular momentum~\eqref{balanceJ}. For a binary system modeled by two point masses, and moving on a circular orbit, the GW polarizations~\eqref{polar} become 
\begin{equation}\label{polarcirc}
\left(\begin{array}{l}h_+\\[0.2cm]h_\times
\end{array}\right) = \frac{2G m \nu}{c^2 d}\left(\frac{G m \omega}{c^3}\right)^{2/3}
 \left(\begin{array}{l} \bigl(1 +\cos^2\!i\bigr)\cos(2\phi) \\[0.2cm]\bigl(2 \cos i\bigr) \sin(2\phi)
\end{array}\right) 
\,,
\end{equation}
where $\nu=m_1 m_2/(m_1+m_2)^2$ denotes the symmetric mass ratio between the two compact objects, $m=m_1+m_2$ is the total mass, and $i$ is the inclination angle of the binary's orbital plane with respect to the plane of the sky (the polarization vector $\bm{p}$ pointing by convention towards the ``ascending node''). In the quadrupole approximation the phase of the signal is $\phi^\text{GW}=2\phi$, where $\phi = \int \omega \,\ud t$ is the orbital phase and $\omega$ the angular frequency; the signal frequency is usually denoted $f^\text{GW}=\omega/\pi$. 

For the purposes of detection and subsequent data analysis, the most important information provided by GR is the time evolution of the phase and frequency, that are computed from the energy balance prescription~\eqref{balanceE}. For circular orbits neither the angular momentum balance equation~\eqref{balanceJ} nor the averaging procedure are necessary. Both $E$ and the flux $F^\text{GW}$ are only functions of the orbital frequency $\omega$, and the flux is readily computed from Eq.~\eqref{fluxE} in the case of a Newtonian system of two point masses on a circular orbit. Hence the balance equation becomes an ordinary differential equation for the frequency:
\begin{equation}\label{chirp}
\frac{\dot\omega}{\omega^2}  = \frac{96\nu}{5} \left(\frac{G m \omega}{c^3}\right)^{5/3}\,.
\end{equation}
By integrating this equation one successively obtains the orbital frequency as a function of time, and the orbital phase as a function of frequency. For convenience we use the dimensionless variables
\begin{equation}\label{dimensionless}
x = \left(\frac{G m \omega}{c^3}\right)^{2/3}\,,\qquad \Theta = \frac{\nu c^3}{5G m}\bigl(t_\text{c}-t\bigr)\,.
\end{equation}
Note that $x$ can be seen as a small PN parameter of the order of $\mathcal{O}(\epsilon_\text{PN}^2)$; and $t_\text{c}$ denotes the instant of coalescence, at which the distance between the particles formally vanishes and the frequency diverges. With those notations, and with $\phi_0$ denoting an initial constant phase, we find
\begin{subequations}\label{xphisol}
\begin{align}
x &= \frac{1}{4}\Theta^{-1/4}\,,\label{xsol}\\
\phi &= \phi_0  - \frac{x^{-5/2}}{32\nu}\,.\label{phisol}
\end{align}
\end{subequations}
These formulas, together with~\eqref{polarcirc}, describe the ``chirp'' of GWs at the lowest approximation, \textit{i.e.}, the way the frequency, phase and amplitude of the signal increase, untill some point at the onset of the merger of the compact objects, at which the approximation is no longer valid.

Inspection of Eqs.~\eqref{polarcirc}--\eqref{xphisol} shows that the GW signal depends on one combination of the two masses only, called the chirp mass and given by $\mathcal{M} = m \,\nu^{3/5}$. Roughly speaking there are two observables, the amplitude $h$ and the frequency chirp $\dot{\omega}$, from which one can determine at once the chirp mass $\mathcal{M}$ and the distance $d$. Actually things are more complicated because in a given detector, the measured amplitude is a certain linear combination~\eqref{h} of the two polarizations, which also depends on the direction of the source and its orientation with respect to the detector. We need thus several detectors to determine these extra angles, and also the inclination angle $i$ in Eq.~\eqref{polarcirc}. Note that the measured mass is the redshifted one, $\mathcal{M}=(1+z)\mathcal{M}_\text{source}$, where $z$ is the cosmological redshift of the source ($z\sim 0.1$ in the case of GW150914) and $\mathcal{M}_\text{source}$ is the actual chirp mass of the binary. As for the measured distance, it is exactly the ``luminosity distance'' $d \equiv d_\text{L}$ used by cosmologists, who refer to the chirping binaries as standard sirens. Finally, the quadrupole formula gives consistent estimates (even for GW150914\,!) for the mass, the distance, the maximal amplitude of the signal, and the number of orbital cycles from the entry frequency of the detector's band till the merger, with the \textit{proviso} that we suplement the quadrupole formula with an information from full GR, namely that the merger occurs at a separation of the order of the mass. Then, of course, the merger itself can only be described numerically.

In GR there is a notion of the total energy contained in the space-time, including both the contribution from the matter sources and that of the gravitational field: This is the so-called Arnowitt-Deser-Misner (ADM) mass-energy~\cite{ADM}, which is exactly conserved. Let us apply this notion to the problem of the coalescence of two black holes, assuming a very crude model, in which the binary's orbit is merely Newtonian and circular, and the radiation field is described by the quadrupole formula~\eqref{balanceE}. At very early times $t\to-\infty$, the black holes were moving almost freely on quasi-hyperbolic orbits, and later formed a gravitationally bound system by GW emission, which then spiralled in till the merger. At any time the ADM energy is 
\begin{equation}\label{EADM}
E_\text{ADM} = m c^2 - \frac{G m^2 \nu}{2r} + \int_{-\infty}^t \ud t' \,F^\text{GW}(t') \,,
\end{equation}
where the quadrupole flux is given by~\eqref{fluxE} --- this equation is an integrated version of Eq.~\eqref{balanceE}, which represented
the time-variation of the Bondi mass~\cite{BBM62}. Since initially the black holes were infinitely far apart ($r\to\infty$), the conserved ADM energy is just equal to $m c^2$, while after the merger it also equals
\begin{equation}\label{EADMc}
E_\text{ADM} = M_\text{c} c^2 + \int_{-\infty}^{t_\text{c}} \ud t \,F^\text{GW}(t) \,,
\end{equation}
where $M_\text{c}$ is the mass of the black hole formed by the coalescence, occuring at the instant $t_\text{c}$. Hence we see, using the constancy of the ADM mass, that the energy which has been radiated away during the process is
\begin{equation}\label{DeltaEGW}
\Delta E^\text{GW} = (m - M_\text{c}) c^2 = \int_{-\infty}^{t_\text{c}} \ud t \,F^\text{GW}(t) =  \frac{G m^2 \nu}{2r_\text{c}} \,.
\end{equation}
That is, it is equal to the mechanical binding energy of the binary system at the instant of coalescence $t_\text{c}$, which occurs when the separation is of the order of the mass, say $r_\text{c}\sim\frac{2 G m}{c^2}$. In the case of GW150914 the final mass $M_\text{c}$ has been measured, and about three solar masses have been released in the form of GWs in a few tens of second. This corresponds to a total power of the order of $10^{49}\,\text{W}$, well consistent with the result of sophisticated numerical simulations of the merger of two black holes in GR. This power is huge, but notice that it is only about a thousandth of $c^5/G$, which represents the natural general relativistic scale for a power, \textit{i.e.}, the Planck scale which happens in this case not to depend on Planck's constant $\hbar$. Thus we may imagine that the gravitational astronomy could some day bring us a surprise, with the discovery of a GW source that is even more powerful than a binary black hole system.

\section{Analytic approximation methods for computing the GW chirp}
\label{sec:analapprox}

Though the quadrupole formula is very useful, it is not sufficient when we want to perform precise calculations, and it becomes inoperational in the final merger phase of the compact objects. As we have seen, it does not permit to measure the two masses $m_1$ and $m_2$ separately, but only the chirp mass $\mathcal{M}$. The degeneracy over the masses is removed by including the relativistic PN corrections $\sim \mathcal{O}(\epsilon_\text{PN}^2)$ in the quadrupole formula~\eqref{balanceE}. In addition, the PN corrections depend on the spins of the compact objects (\textit{i.e.}, their intrinsic classical angular momenta). Taking into account the spins in the GR templates is important, and their measurement represents a valuable astrophysical information.

At the beginning of the building of the LIGO/Virgo detectors in the end of the 1980s, it was thought that the quadrupole formula is sufficient for detecting and analyzing binary neutron star coalescences. But in the early 1990s it was realized~\cite{3mn,BS93} that neutron star binaries spend thousands of cycles in the band of the detectors, and that the GW templates should be able to monitor the signal with a precision of a fraction of a cycle over the entire bandwidth, say $\frac{\delta\phi}{2\pi}\lesssim 0.1$ where $\phi$ is the orbital phase~\eqref{phisol}. It was then estimated that PN corrections in the quadrupole formula must be developed up to at least the daunting 3PN order $\sim\epsilon_\text{PN}^6$. Also recall that the quadrupole formula, as seen as a small radiation reaction contribution in the dynamics of the source, is itself a 2.5PN effect $\sim\epsilon_\text{PN}^5$ relatively to the Newtonian acceleration. This shows the highly relativistic character of compact binary systems observed in the LIGO/Virgo band as compared to the binary pulsar, for which the PN corrections in the orbital $\dot{P}$ are negligible. At the time of the binary pulsar only the 1PN correction to the quadrupole formula was known~\cite{EW75,WagW76,BD89,BS89}. Furthermore, it was also realized that the PN corrections are important not only for the precise off-line analysis of the signals once they are detected, but also for the on-line process of detection~\cite{3mn}. 

The coalescence signal of two compact objects can be decomposed into three phases, see Fig.~\ref{fig2}:
\begin{figure}[t]
\begin{center}
\includegraphics[width=13cm]{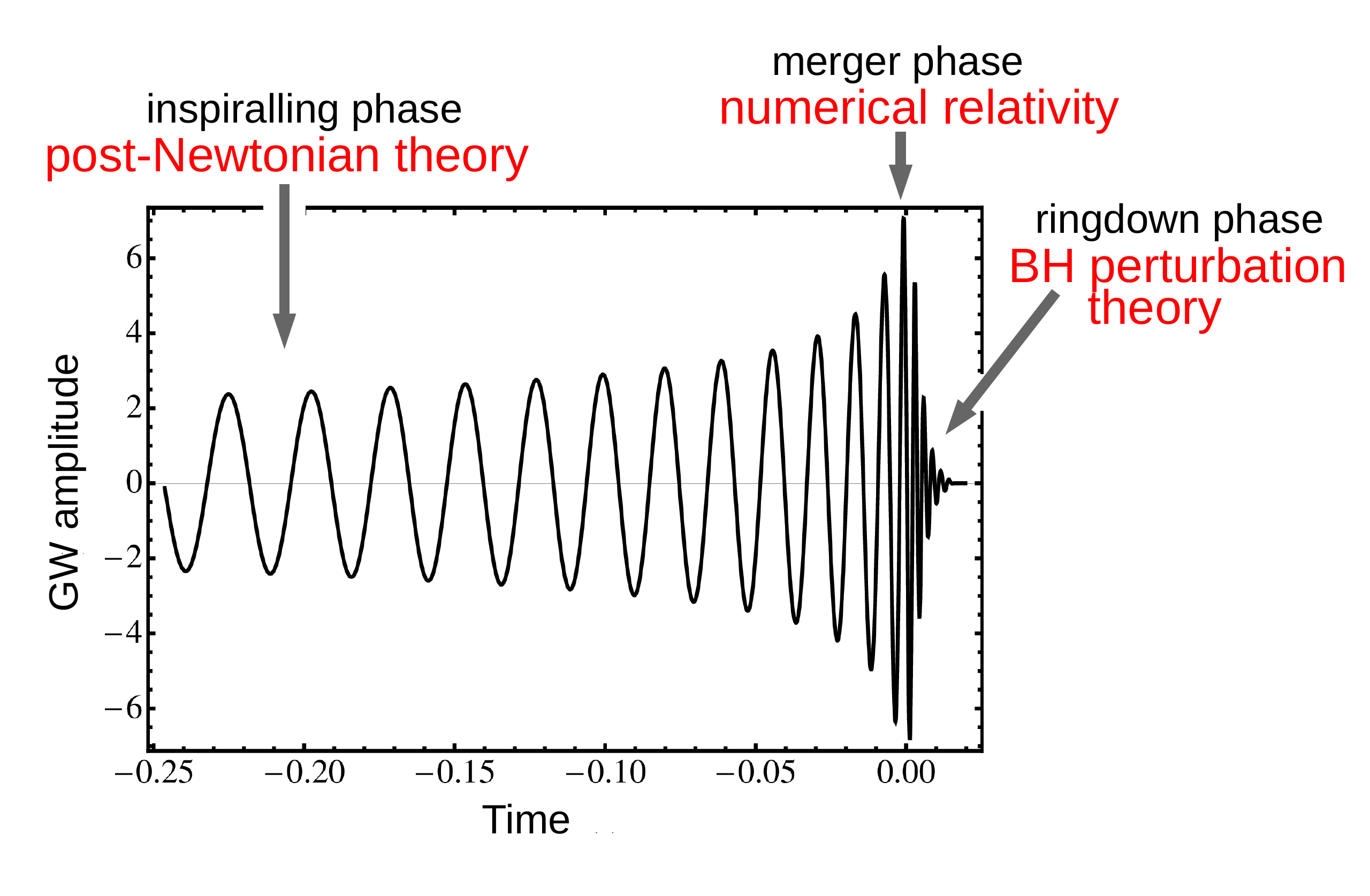}
\end{center}
\caption{The three phases of the coalescence of two compact objects, with the theoretical methods required to build accurate templates in each phases. While neutron star binaries are observed in the band of the LIGO-Virgo detectors mostly during the early inspiral phase, the more massive black hole binaries are essentially seen in the last orbits of the inspiral and in the final merger and ringdown phases.}\label{fig2}
\end{figure}
\begin{itemize}
\item The early inspiral during which the frequency and amplitude of the signal chirp with time (the chirp was discussed and defined in Sec.~\ref{sec:gravastro}). During this phase the signal is universal, \textit{i.e.}, it does not depend on the nature of the compact objects, be they black holes or neutron stars or more exotic objects like boson stars. The signal depends only on the masses and the spins. The PN approximation constitutes the ideal tool for describing the inspiral phase. For low mass compact binaries such as double neutron stars, the detectors are mostly sensitive to the inspiral phase, and the currently known analytical PN templates are accurate enough for detection and analysis, at least for moderate spins. Thus, the data analysis of neutron star binary events like GW170817~\cite{GW170817} is based on the PN templates.
\item The merger phase, when the dynamics undergoes a transition from adiabatic inspiral to some unstable plunge, followed by the rapid collapse of the two objects to form a black hole. Numerical relativity (NR) has succeeded in the 2005s to compute the merger of two black holes~\cite{Pret05,Bak06,Camp06}. A non trivial point (which did not seem to be obvious some years ago~\cite{BCT98}), is that the overlap between the PN and NR regimes exists and is quite significant.\footnote{Because of prohibitive computing times the NR calculations are limited to a few tens of orbits before the merger, and will likely never be competitive with the PN approximation when monitoring tens of thousands of cycles in the early inspiral.} The important issue of matching the PN and NR waveforms has been solved using several techniques~\cite{Boyle08,Pan10}. One is the hybrid inspiral-merger-ringdown (IMR or IMR-Phenom), which consists of introducing between the PN and NR domains of validity an overlapping time interval that is parametrized in a phenomenological way~\cite{Ajith10,Ajith11}. The other technique recasts the actual PN two-body equations of motion and radiation into a simpler effective one-body (EOB) form~\cite{BuonD99,DNorleans}. The EOB dynamics is described in a non perturbative way, which permits to extend the domain of validity of the PN approximation. A variant of EOB called EOB-NR is matched to the NR results. Both the IMR and EOB waveforms are extensively used in the LIGO/Virgo data analysis of the recent binary black hole events. 
\item The ``ringdown'', when the newly formed black hole, which is highly deformed due to the nonlinear dynamics of the collision, relaxes to a stationary configuration given by the Kerr solution --- the unique stationary rotating black hole in GR, depending only on the mass and the spin. The perturbed black hole emits quasi-normal mode radiation, and the modes can be analyzed by comparing with black hole perturbation theory~\cite{BCW06}. A test of the ``no-hair'' theorem for black holes can be implemented by looking at the presence of an abnormal quadrupole moment endowed by the black hole, which would be independent from the mass and the spin. Alternatively this can be viewed as a test of the existence of a new, exotic form of matter alternative to black holes. 
\end{itemize}

\begin{figure}[t]
	\begin{center}
		\begin{tabular}{c}
		\hspace{-0.9cm}\includegraphics[width=8.5cm,angle=0]{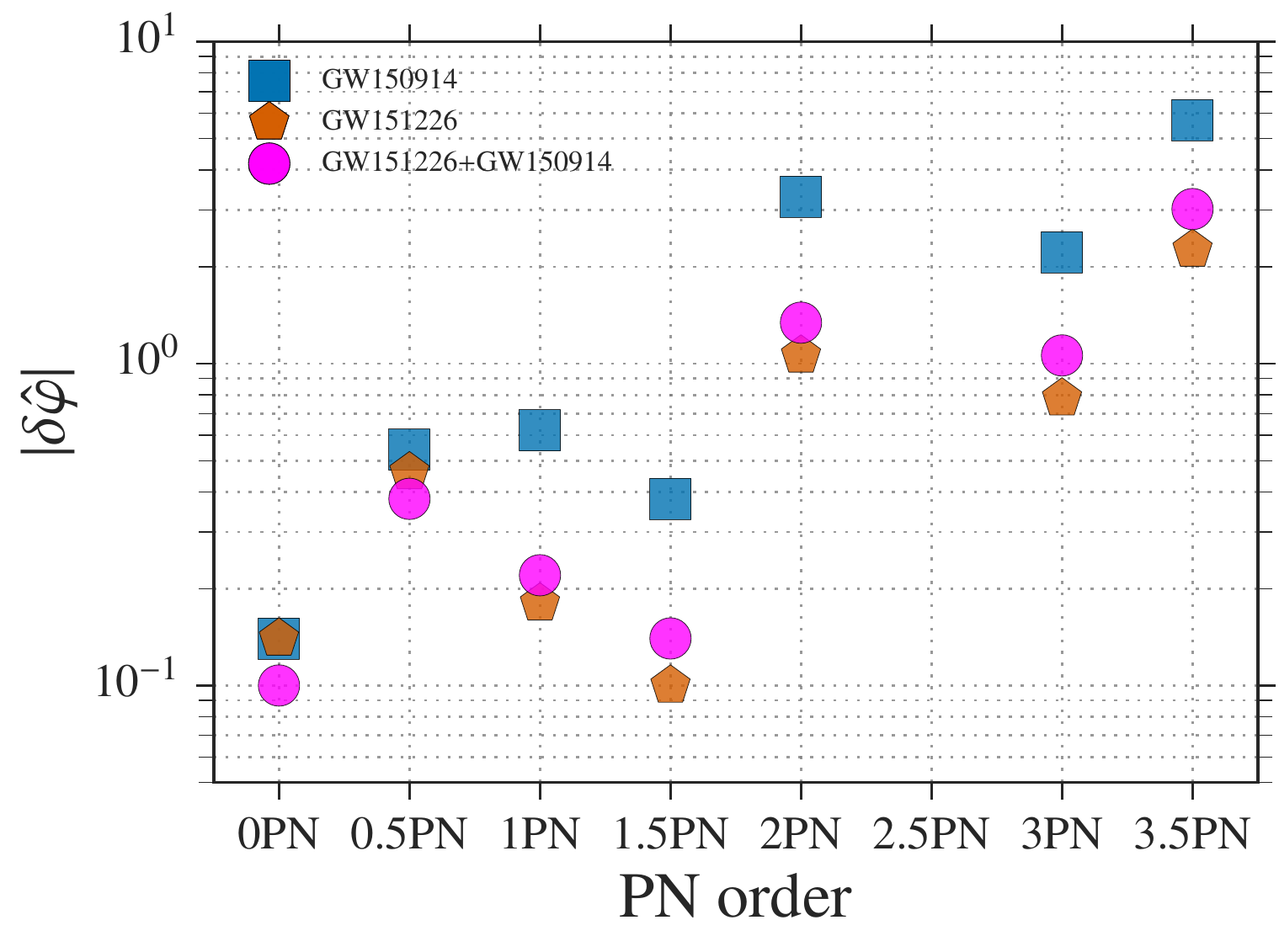}
		\hspace{0.3cm}\includegraphics[width=8.4cm,angle=0]{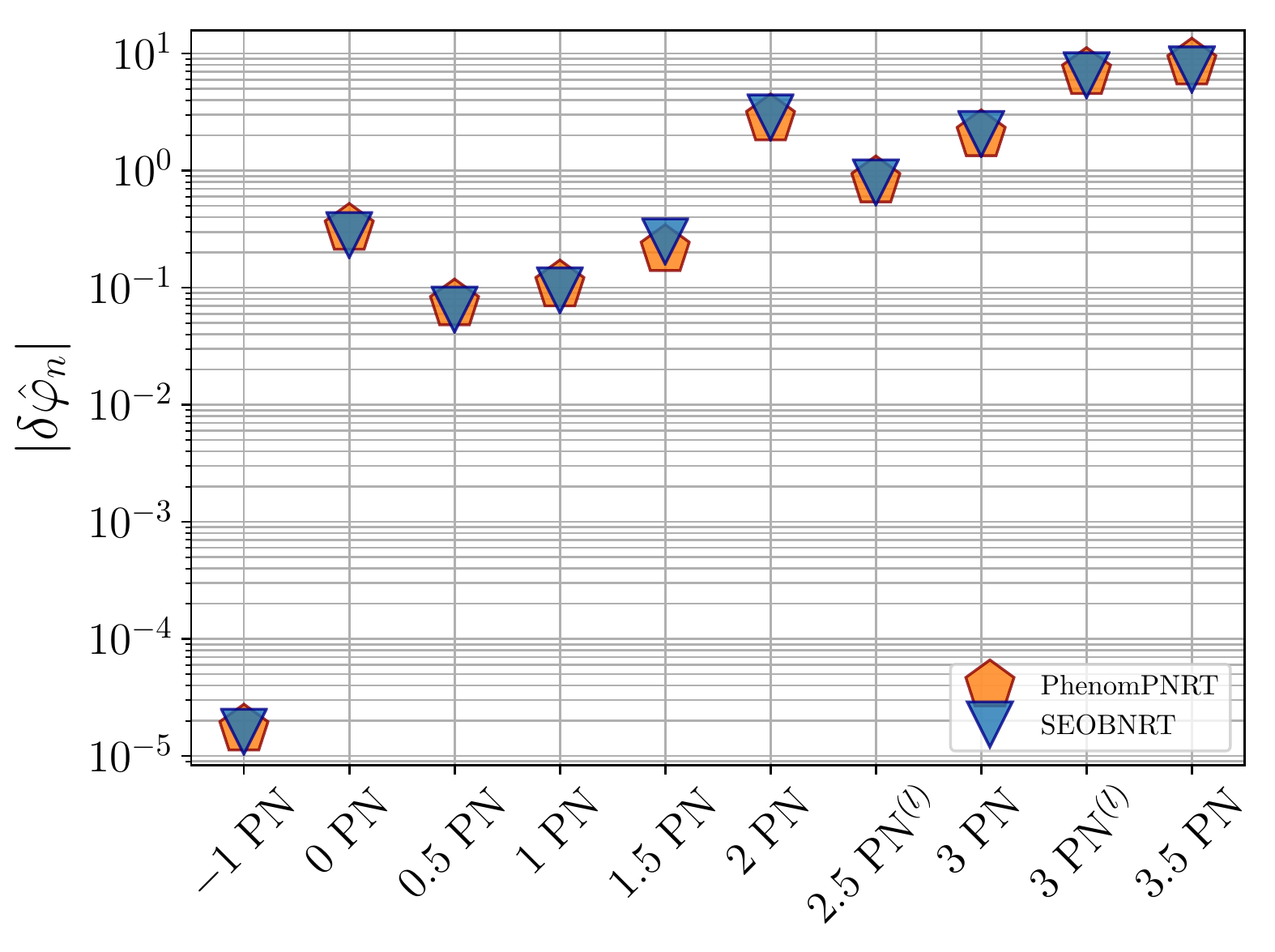}
		\end{tabular}
		\caption{Observational constraints on the PN parameters, from measurements of the black hole events GW150914 and GW151226 (left panel) and from the neutron star event GW170817 (right panel). The limits are obtained by assuming the GR values~\eqref{PNparameters} for all the PN parameters but for one. This particular one is allowed to vary and is measured by the technique of matched filtering. For instance the 1.5PN parameter agrees with the GR prediction $\varphi_\text{1.5PN}=-10\pi$ within a fractional accuracy of the order of 10 \%, which constitutes an interesting test of the tail effect~\cite{BS93,BSat95}.}
		\label{fig3}
	\end{center}
\end{figure}

We now review the state-of-the-art on PN approximations applied to the inspiral of two compact objects. The first problem is that of the equations of motion, and has been solved up to the 4PN order $\sim\epsilon_\text{PN}^8$ for non-spinning compact bodies.\,\footnote{See~\cite{BlanchetLR} for an history of the problems of motion and radiation, and for references to previous PN approximations.} Different methods have been used, with equivalent results: The Hamiltonian formalism in ADM coordinates~\cite{JaraS12,JaraS13,DJS14}, and the Fokker action of GR in harmonic coordinates~\cite{BBBFMa,BBBFMb,BBBFMc,MBBF17}. In addition partial results have been obtained with the effective field theory~\cite{FS4PN,GLPR16}. The second problem is the one of the GW field, and of course, this is that problem whose solution is directly used by LIGO/Virgo. Here the state-of-the-art is 3.5PN order $\sim\epsilon_\text{PN}^7$ beyond the result of the quadrupole formula~\cite{BS89,BS93,B95,BDIWW95,B98tail,BIJ02,BFIJ02,BI04mult,BDEI04}, and the 4.5PN term is also known~\cite{MBF16}. To reach this result a cocktail of approximation methods in GR called ``MPM-PN'' has been used:
\begin{itemize}
\item In a first stage we control the gravitational field generated by an isolated matter system in the exterior zone of the system. A non-linearity or \textit{post-Minkowskian} (PM) expansion is combined with a multipolar (M) expansion parametrized by some sets of source multipole moments, yielding the most general solution of the Einstein field equation in the exterior zone~\cite{BD86,B87,BD92,BDI95}. In particular this solution recovers the Bondi-Sachs formalism~\cite{BBM62,Sachs62} for the asymptotic structure of radiative fields at infinity from the matter source. This is the MPM part of the method.
\item The MPM solution is matched to the PN field in the near and interior zones of the source. This is achieved by a matching equation, within a specific variant of the theory of matched asymptotic expansions. The matching is performed up to any PN order and yields unique expressions for the multipole moments of the source~\cite{B95,B98mult} as well as for the radiation reaction contributions in the inner PN metric~\cite{PB02,BFN05}. This completes the MPM-PN approach.
\item The MPM-PN solution is applied to systems of compact bodies treated as point masses (possibly with spins), using delta functions singularities. The model thus entails ultra-violet (UV) divergences, that are cured by means of dimensional regularization. Here we borrow dimensional regularization from quantum field theory and use it as a powerful regularisation scheme in classical GR. In addition, independently of the model of point particles, there are infra-red (IR) divergences in the general formalism, treated by a variant of dimensional regularization.
\end{itemize}
We report the most complete results concerning the PN corrections in the orbital phase --- crucial for both processes of detection and subsequent parameter analysis. Extending the ``Newtonian'' result~\eqref{phisol} we have:
\begin{equation}\label{phasePN}
\phi = \phi_0 - \frac{x^{-5/2}}{32\nu} \sum_{p} \Bigl(\varphi_\text{$p$PN} + \varphi_\text{$p$PN}^{(l)} \ln x\Bigr) \,x^p \,,
\end{equation}
where the sum runs over the successive PN corrections: $x^p \sim \mathcal{O}(\epsilon_\text{PN}^{2p})$, with $p$ being an integer or half integer. Some of the PN terms involve the logarithm of $x$, which we indicate by adding a superscript $(l)$ to the PN parameter. Up to the 3.5PN order we have~\cite{BDIWW95,B98tail,BIJ02,BFIJ02,BI04mult,BDEI04}\,\footnote{Here $\gamma_\text{E}$ is the (probably irrational/transcendental) Euler's constant.}
\begin{subequations}\label{PNparameters}
\begin{align}
\varphi_\text{$-$1PN} &= 0\,,\\
\varphi_\text{0PN} &= 1\,,\\
\varphi_\text{0.5PN} &= 0\,,\\
\varphi_\text{1PN} &= \frac{3715}{1008} + \frac{55}{12}\nu\,,\\
\varphi_\text{1.5PN} &= - 10\pi\,,\\
\varphi_\text{2PN} &= \frac{15293365}{1016064} +
  \frac{27145}{1008} \nu + \frac{3085}{144} \nu^2\,,\\
\varphi_\text{2.5PN}^{(l)} &= \left(\frac{38645}{1344} - \frac{65}{16}\nu\right) \pi\,,\\
\varphi_\text{3PN} &= \frac{12348611926451}{18776862720} - \frac{160}{3}\pi^2 -
    \frac{1712}{21}\gamma_\text{E} - \frac{3424}{21} \ln 2 \nonumber\\ &+
    \left(-\frac{15737765635}{12192768} + \frac{2255}{48}\pi^2
    \right)\nu + \frac{76055}{6912}\nu^2 -
    \frac{127825}{5184}\nu^3\,,\\
\varphi_\text{3PN}^{(l)} &= - \frac{856}{21}\,,\\
\varphi_\text{3.5PN} &= \left(\frac{77096675}{2032128} + \frac{378515}{12096}\nu -
  \frac{74045}{6048}\nu^2\right) \pi\,.
\end{align}
\end{subequations}
The PN parameters have been factorized by the dominant quadrupolar effect, so that $\varphi_\text{0PN} = 1$ by definition. Notice the term $\varphi_\text{$-$1PN}$ which corresponds to a dipolar effect; this term would appear in scalar-tensor theory but is absent in GR, hence $\varphi_\text{$-$1PN} = 0$. The 1.5PN term is especially interesting as it corresponds to the dominant nonlinear \textit{tail} effect --- backscattering of linear quadrupolar GWs onto the space-time curvature generated by the mass of the source. In scalar-tensor theory there would be a dipolar tail term at the 0.5PN order~\cite{Laura1,Laura2}, but again this effect is absent in GR. The observational limits on the measurement of the PN parameters by LIGO/Virgo are shown in Fig.~\ref{fig3}. 

Going beyond 3.5PN order, the 4PN parameter has not yet been computed by PN theory, but its leading term in the small mass ratio limit $\nu\to 0$ is known from black hole perturbation theory~\cite{TSasa94,TTS96}:
\begin{subequations}\label{param4PN}
\begin{align}
\varphi_\text{4PN} &= \frac{2550713843998885153}{2214468081745920} - \frac{45245}{756}\pi^2 -
    \frac{9203}{126}\gamma_\text{E} - \frac{252755}{2646} \ln 2 - \frac{78975}{1568} \ln 3 + \mathcal{O}(\nu) \,,\\
\varphi_\text{4PN}^{(l)} &= - \frac{9203}{252} + \mathcal{O}(\nu)\,.
\end{align}
\end{subequations}
We expect that PN theory will be able to fully confirm this result, as well as of course, to provide all the mass ratio corrections $\mathcal{O}(\nu)$ therein. Finally the complete 4.5PN parameter has been derived by PN theory, and is due to an iterated nonlinear tail effect~\cite{MBF16}:
\begin{subequations}\label{param4.5PN}
\begin{align}
\varphi_\text{4.5PN} &= \left( - \frac{93098188434443}{150214901760} + \frac{80}{3}\pi^2 +
    \frac{1712}{21}\gamma_\text{E} + \frac{3424}{21} \ln 2 \right.\nonumber\\ &\left. + \left[\frac{1492917260735}{1072963584} - \frac{2255}{48}\pi^2\right]\nu -
  \frac{45293335}{1016064}\nu^2 -
  \frac{10323755}{1596672}\nu^3\right) \pi \,,\\
\varphi_\text{4.5PN}^{(l)} &= \frac{856}{21} \,\pi \,.
\end{align}
\end{subequations}

We emphasized that in the matched filtering analysis of GWs, it is important to take into account the effects of spins. Here we report the spin-orbit (SO) coupling contributions, which are linear in the two spins and result from the coupling with the orbital angular momentum. They are known up to the 4PN order~\cite{BMB13,MBBB13}:
\begin{subequations}
\begin{align}
\varphi^\text{SO}_\text{1.5PN} &= \frac{1}{G m^2}\left[\frac{235}{6}S_z +\frac{125}{8} \frac{\delta m}{m} \,\Sigma_z\right]\,,\\
\varphi^{\text{SO}\,(l)}_\text{2.5PN} &= \frac{1}{G m^2}\left[\left(-\frac{554345}{2016}-\frac{55}{8}\nu\right)S_z
  +\left(-\frac{41745}{448}+\frac{15}{8}\nu\right) \frac{\delta m}{m} \,\Sigma_z\right]\,,\\
\varphi^\text{SO}_\text{3PN} &= \frac{\pi}{G m^2}\left[\frac{940}{3}\,S_z +\frac{745}{6}\,\frac{\delta m}{m} \,\Sigma_z\right]\,,\\
\varphi^\text{SO}_\text{3.5PN} &= \frac{1}{G m^2}\left[\left(-\frac{8980424995}{6096384}+\frac{6586595}{6048}\nu
  -\frac{305}{288}\nu^2\right)S_z\right.
  \nonumber\\ &\qquad\qquad\qquad\left. +\left(-\frac{170978035}{387072}
  +\frac{2876425}{5376}\nu+\frac{4735}{1152}\nu^2\right) \frac{\delta m}{m} \,\Sigma_z\right]\,,\\
\varphi^\text{SO}_\text{4PN} &= \frac{\pi}{G m^2}\left[\left(
  \frac{2388425}{3024} - \frac{9925}{36}\nu \right) S_z +
  \left( \frac{3237995}{12096} - \frac{258245}{2016}\nu
  \right)\frac{\delta m}{m} \,\Sigma_z\right]\,,
\end{align}
\end{subequations}
where the mass difference is denoted by $\delta m = m_1-m_2$, $\bm{S}_1$ and $\bm{S}_2$ are the two individual spins, and $S_z$ and $\Sigma_z$ are the projections of the particular combinations $\bm{S}=\bm{S}_1+\bm{S}_2$ and $\bm{\Sigma}=\frac{m}{m_2}\bm{S}_2-\frac{m}{m_1}\bm{S}_1$ perpendicular to the orbital plane, \textit{i.e.}, parallel to the orbital angular momentum. The quadratic spin-spin (SS) coupling contributions to the orbital phase are also known~\cite{BFMP15}.

\section{Influence of the internal structure of compact bodies}
\label{sec:finitesize}

The PN parameters have been obtained within the so-called ``pole-dipole'' model, which approximates the rotating compact body as a point mass with a spin, but neglects the effect of the finite size and the internal structure of the body, such as the internal velocity field and the type of equation of state. In particular, the quadrupolar tidal deformation of the body is ignored. Neutron stars have a strong internal gravity so it is very difficult to deform them. We expect that they should be distorted by the gravitational field of the companion only in the last orbits before the merger~\cite{Kochanek92,BCutler92}. On the other hand, the numerical computation of the merger of two neutron stars shows that it is strongly dependent on the internal structure and the (unknown) equation of state~\cite{ShibataLR,FaberRasioLR}. Therefore a legitimate question to ask, is whether and at which PN order the internal structure of \textit{extended} compact objects influences the orbital phase evolution. Here we answer this question by means of a simple Newtonian model for the tidal interaction between extended bodies without spins during the inspiral phase, at the lowest quadrupolar level.\,\footnote{See Refs.~\cite{FHind08,Hind08,DN09tidal,BP09,BiniDF12,F14} for entries in the literature.} The Newtonian equations of motion of $N$ extended spinless bodies ($a,b=1,\dots, N$), to linear order in the quadrupole moments, are
\begin{equation}\label{EOM}
m_a \frac{\ud v_a^i}{\ud t} = G \sum_{b \not= a} \left[m_a m_b \frac{\partial}{\partial y_a^i}\left(\frac{1}{r_{ab}}\right) + \frac{1}{2}\left( m_a \,q_b^{jk} + m_b \,q_a^{jk} \right)\frac{\partial^3}{\partial y_a^i\partial y_a^j\partial y_a^k}\left(\frac{1}{r_{ab}}\right)\right] \,,
\end{equation}
where $m_a$ are the masses, and we denote the position and velocity of the center of mass of the bodies by $\bm{y}_a(t)$ and $\bm{v}_a(t)=\ud \bm{y}_a/\ud t$, with the Euclidean separation between centers of mass being $r_{ab}=\vert \bm{y}_a-\bm{y}_a\vert$. The quadrupole moments of the bodies, supposed to be made of a perfect fluid, read
\begin{equation}\label{qaij}
q_a^{ij} = \int_{\mathrsfs{V}_a} \ud^3\bm{z}_a \,\rho_a\Bigl( z_a^i z_a^j - \frac{1}{3} \delta^{ij} \bm{z}_a^2 \Bigr)\,,
\end{equation}
with $\mathrsfs{V}_a$ the volume of the body, $\bm{z}_a=\mathbf{x}-\bm{y}_a(t)$ the distance between a generic point $\mathbf{x}$ inside the body and the center of mass, $\rho_a=\rho(\bm{y}_a+\bm{z}_a,t)$ the Newtonian mass density of the body, $\rho(\mathbf{x},t)$ being the usual Eulerian density. The mass-centred condition reads
\begin{equation}\label{CM}
\int_{\mathrsfs{V}_a} \ud^3\bm{z}_a \,\rho_a\,z_a^i = 0\,.
\end{equation}
The conserved energy of the $N$-body system is the sum of the internal (Newtonian) energies $e_a$ and of the orbital contributions, including the quadrupole effects:
\begin{equation}\label{E}
E = \sum_a \biggl\{ e_a + \frac{1}{2}m_a \bm{v}_a^2 - \frac{G}{2} \sum_{b \not= a} \frac{m_a m_b}{r_{ab}} - \frac{1}{2} \,q_a^{ij} \mathcal{E}_a^{ij}\biggr\}\,,
\end{equation}
where we have introduced the tidal field acting on body $a$ and due to the other bodies $b \not= a$:
\begin{equation}\label{tidalfield}
\mathcal{E}_a^{ij} \equiv  G \sum_{b\not= a} m_b \frac{\partial^2}{\partial y_a^i\partial y_a^j}\!\left(\frac{1}{r_{ab}}\right)\,.
\end{equation}
Posing $\bm{w}_a=\ud \bm{z}_a/\ud t$ for the internal velocity field of body $a$, $\Pi_a=\Pi(\bm{y}_a+\bm{z}_a,t)$ for the specific internal energy satisfying the thermodynamical relation $\ud\Pi=- P\,\ud(1/\rho)$ (with $P$ the pressure), and $u_a$ for the internal self-gravity given by the Poisson integral over the volume of the body, we have
\begin{equation}\label{ea}
e_a = \int_{\mathrsfs{V}_a} \ud^3\bm{z}_a \,\rho_a\left( \frac{1}{2}\bm{w}_a^2 + \Pi_a - \frac{u_a}{2} \right)\,.
\end{equation}
The coupling of the quadrupole moment $q_a^{ij}$ with the external tidal field $\mathcal{E}_a^{ij}$ of the other bodies implies a variation of the internal energy given by
\begin{equation}\label{dedt}
\frac{\ud e_a}{\ud t} = \frac{1}{2} \frac{\ud q_a^{ij}}{\ud t} \mathcal{E}_a^{ij} \,.
\end{equation}

We consider the case where the quadrupole moment is induced by the tidal field of the other bodies. To linear order, we can introduce a coefficient $\lambda_a$ characterizing the deformability (or ``polarizability'') of the body under the influence of the external field, such that
\begin{equation}\label{lambda}
q_a^{ij} = \lambda_a \,\mathcal{E}_a^{ij}\,.
\end{equation}
The ``response'' coefficient $\lambda_a$ depends on the internal structure of the body, and is commonly given as $\lambda_a=\frac{2}{3G} \,k_a r_a^5$ in terms of the radius $r_a$ of the body and the mass-type quadrupolar Love number $k_a \equiv k_a^{(2)}$ (see for instance~\cite{Hind08}). In fact it will be more convenient to characterize the internal structure of the body by the dimensionless parameter
\begin{equation}\label{Lambda}
\Lambda_a = \frac{c^{10}}{G^4 m_a^5}\lambda_a = \frac{2}{3} k_a \left(\frac{c^2 r_a}{G m_a}\right)^5\,.
\end{equation}
In the case of the induced quadrupole moments~\eqref{lambda}, the total energy of the system becomes
\begin{equation}\label{Einduced}
E = \sum_a \biggl\{ \frac{1}{2}m_a \bm{v}_a^2 - \frac{G}{2} \sum_{b \not= a} \frac{m_a m_b}{r_{ab}} - \frac{1}{4} \lambda_a \,\mathcal{E}_a^{ij} \mathcal{E}_a^{ij}\biggr\}\,.
\end{equation}

Consider a compact binary system ($N=2$) moving of an exact circular orbit. From Eq.~\eqref{lambda} we see that the two quadrupole moments face each other, and remain constant along the circular orbit. The equation of the relative motion reduces to $\ud\bm{v}/\ud t=-\omega^2 \bm{x}$, where $\bm{x}=\bm{y}_1-\bm{y}_2$ and $\bm{v}=\ud \bm{x}/\ud t$ are the relative position and velocity (with $r \equiv r_{12}$). We find from~\eqref{EOM} the orbital frequency
\begin{equation}\label{omega}
\omega^2 = \frac{G m}{r^3} \biggl[ 1 + 9 \nu \bigl( X_1^3 \Lambda_1 + X_2^3 \Lambda_2\bigr) \gamma^5\biggr]\,.
\end{equation}
We pose $X_a=m_a/m$ so that $X_1 X_2=\nu$ is the symmetric mass ratio, denote $\gamma=\frac{G m}{r c^2}$ and employ the notation~\eqref{Lambda}. In turn the conserved energy~\eqref{Einduced} for circular orbits reduces to
\begin{equation}\label{Ecirc}
E = - \frac{G m^2 \nu}{2 r}\biggl[ 1 - 6 \nu \bigl( X_1^3 \Lambda_1 + X_2^3 \Lambda_2\bigr) \gamma^5\biggr] \,.
\end{equation}
As the effect of the deformation of the bodies computed here is purely ``Newtonian'', we see that the $c$'s we have introduced into our definitions naturally cancel out in Eqs.~\eqref{omega} and~\eqref{Ecirc}.

The above dynamics is \textit{conservative}, \textit{i.e.}, we have neglected the dissipative radiation reaction effect on the orbit. This effect is taken into account when we impose the flux-balance equation~\eqref{balanceE}. Again there is no need to impose the angular momentum balance equation~\eqref{balanceJ} for circular orbits. The total quadrupole moment of the system is the sum of the orbital one and of the intrinsic moments of the bodies, given by~\eqref{qaij}: 
\begin{equation}\label{Qijtot}
Q^{ij} = m \nu \Bigl( x^i x^j - \frac{1}{3} \delta^{ij} r^2\Bigr) + q_1^{ij} + q_2^{ij} \,.
\end{equation}
Plugging this into the flux formula, computing the time derivatives using the equations of motion including the contributions from the quadrupole moments, see Eq.~\eqref{omega}, and keeping only the terms linear in these quadrupoles, yields the flux (still for exact circular orbits) as
\begin{equation}\label{fluxr}
F^\text{GW} = \frac{32 G}{5 c^5} r^4 \omega^6 m^2 \nu^2 \biggl[ 1 + 6 \bigl( X_1^4 \Lambda_1 + X_2^4 \Lambda_2 \bigr) \gamma^5 \biggr]\,.
\end{equation}

\begin{figure}[t]
\begin{center}
\includegraphics[width=10cm]{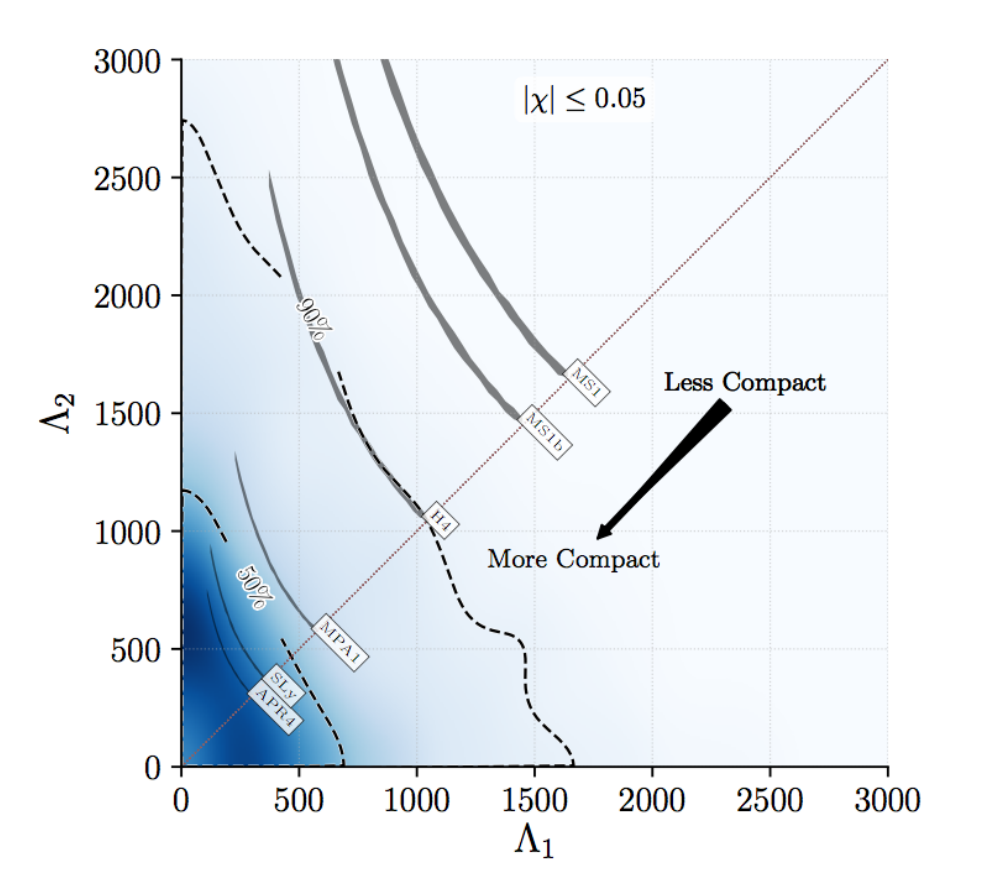}
\end{center}
\caption{Observational constraints on the tidal deformability (or polarizability) and the inner equation of state of neutron stars obtained with GW170817~\cite{GW170817}. The parameters $\Lambda_a$ are defined by~\eqref{Lambda}. Contours enclosing 90\% and 50\% of the probability density are shown with dashed lines. The predictions for tidal deformability given by a set of representative equations of state are given with grey lines. For a stiff equation of state the pressure increases a lot for a given increase in density (for instance $P\propto\rho^\gamma$ with a large value of the polytropic index $\gamma$), thus it gives more resistance to the gravitational force and the neutron star is less compact. The stiffest equations of state are excluded, while the softest (which predict more compact neutron stars) are still allowed; they appear in the dark blue region. The constraints are shown for a low-spin scenario, with dimensionless spin parameter $\vert\chi\vert\leqslant 0.05$, which is probably favored for neutron stars.}\label{fig4}
\end{figure}
At this stage we reexpress the invariants $E$ and $F^\text{GW}$ in terms of the orbital frequency $\omega$ instead of the separation distance $r$ using Eq.~\eqref{omega}. The interest of doing this in GR (\textit{e.g.}, when doing relativistic PN calculations), comes from the fact that the separation $r$ depends on the choice of the coordinate system, while the orbital frequency $\omega$ is invariantly defined in a large class of coordinate systems. Recalling the definition for the invariant dimensionless PN parameter $x$ in~\eqref{dimensionless}, we obtain
\begin{subequations}\label{EFcircx}
\begin{align}
E &= - \frac{1}{2}m \nu c^2 x \biggl[ 1 - 9 \nu \bigl( X_1^3 \Lambda_1 + X_2^3 \Lambda_2\bigr) x^5\biggr]\,,\label{Ecircx}\\
F^\text{GW} &= \frac{32 c^5}{5 G} x^5 \nu^2 \biggl\{ 1 + 6 \Bigl[ \left(X_1+2\nu\right)X_1^3 \Lambda_1 + \left(X_2+2\nu\right)X_2^3 \Lambda_2\Bigr] x^5\biggr\}\,.\label{Fcircx}
\end{align}
\end{subequations}
At this stage we can already draw a firm conclusion: The effect of the internal structure of non-spinning bodies is proportional to $x^5$, and is thus comparable to a relativistic effect occuring at the 5PN order. Recall though that we computed this effect using merely Newton's law of gravity. Of course the latter estimate is just formal, but we expect it to be physically correct in the case of \textit{compact} bodies. But the numerical value of the coefficient involving the $\Lambda_a$'s is to be taken into account. For instance $\frac{G m_a}{c^2 r_a}\sim 0.15$ for neutron stars (hence $x^5\sim 8\,10^{-5}$ at the merger), and the numerical estimates of the Love numbers for neutron stars are of the order of one or say, a tenth~\cite{DN09tidal,BP09}. Therefore the deformability parameters~\eqref{Lambda} for compact bodies should be of the order of $\Lambda_a \sim 1000$, depending of course on the equation of state, as shown in Fig.~\ref{fig4}.

As in Sec.~\ref{sec:quadformula} the phase and frequency evolution follow from~\eqref{balanceE}, where both $E$ and $F^\textit{GW}$ have been computed for the conservative dynamics in Eqs.~\eqref{EFcircx}. This approximation is justified as we are interested in the secular, adiabatic evolution of the orbit over a radiation reaction time scale. We need thus to evaluate the secular variation of the energy $E$, which we immediately find from~\eqref{Ecircx} to be
\begin{equation}\label{dEdx}
\frac{\ud E}{\ud t} = -\frac{1}{2} m \nu c^2 \biggl[ 1 - 54 \nu \Bigl( X_1^3 \Lambda_1 + X_2^3 \Lambda_2 \Bigr) x^5\biggr] \dot{x}\,.
\end{equation}
Combining this with the flux~\eqref{Fcircx}, we get an ordinary differential equation for $x$. It turns out to depend on the following combination of the two deformability parameters:
\begin{equation}\label{Lamdatilde}
\tilde{\Lambda} = \frac{16}{13}\,\biggl[ \bigl( X_1 + 11 \nu\bigr) X_1^3 \Lambda_1 + \bigl( X_2 + 11 \nu\bigr) X_2^3 \Lambda_2 \biggr]\,,
\end{equation}
so normalized that in the case of two identical neutron stars (with the same mass, $X_1=X_2=\frac{1}{2}$, and the same equation of state) it reduces to $\tilde{\Lambda}=\Lambda_1=\Lambda_2$. We obtain the quadrupole finite size effect due to the internal structure on the frequency and phase evolution, extending the point-mass results given by~\eqref{xphisol}, as
\begin{subequations}\label{xphisolintstruct}
\begin{align}
x &= \frac{1}{4}\Theta^{-1/4}\left[ 1 + \frac{39}{8192} \tilde{\Lambda} \,\Theta^{-5/4}\right]\,,\\
\phi &= \phi_0 - \frac{x^{-5/2}}{32\nu} \left[ 1 + \frac{39}{8} \tilde{\Lambda} \,x^{5}\right]\,.
\end{align}
\end{subequations}
These corrections should be added linearly to the purely gravitational PN corrections presented in Sec.~\ref{sec:analapprox}. For compact bodies the effect appears at the very small order 5PN, but for neutron stars becomes relatively large at the merger, of the order of a radian in the phase. Remarkably, it has been possible to put a bound on the tidal deformability of neutron stars with the recent binary neutron star event GW170817, and to infer a constraint on several possible equations of state for the nuclear matter inside neutron stars, see Fig.~\ref{fig4}.


\bibliography{ListeRef_CRAS2019.bib}

\end{document}